# Improving spatial resolution of scanning SQUID microscopy with an on-chip design

Y. P. Pan[1,2], J. J. Zhu[1], Y. Feng[1], Y. S. Lin[1], H. B. Wang[2], X. Y. Liu[2], H. Jin[2], Z. Wang[2], L. Chen[2, *], Y. H. Wang[1, 3*]

1. Department of Physics and State Key Laboratory of Surface Physics, Fudan University, Shanghai 200438, China

2. Center for Excellence in Superconducting Electronics, State Key Laboratory of Functional Material for Informatics, Shanghai Institute of Microsystem and Information Technology, Chinese Academy of Sciences, Shanghai 200050, China

3. Shanghai Research Center for Quantum Sciences, Shanghai 201315, China

* To whom correspondence and requests for materials should be addressed.

Email: leichen@mail.sim.ac.cn ; wangyhv@fudan.edu.cn

**Abstract**

**Scanning superconducting quantum interference device microscopy (sSQUID) is currently one of the most effective methods for direct and sensitive magnetic flux imaging on the mesoscopic scale. A SQUID-on-chip design allows integration of field coils for susceptometry in a gradiometer setup which is very desirable for measuring magnetic responses of quantum matter. However, the spatial resolution of such a design has largely been limited to micrometers due to the difficulty in approaching the sample. Here, we used electron beam lithography technology in the fabrication of the 3D nano-bridge-based SQUID devices to prepare pick-up coils with diameters down to 150 nm. Furthermore, we integrated the deep silicon etching process in order to minimize the distance between the pick-up coil and the wafer edge. Combined with a tuning-fork-based scanning head, the sharpness of the etched chip edge enables a precision of 5 nm in height control. By scanning measurements on niobium chessboard samples using these improved SQUID devices, we demonstrate sub-micron spatial resolutions in both magnetometry and susceptometry, significantly better than our previous generations of nano-SQUIDs. Such improvement in spatial resolution of SQUID-on-chip is a valuable progress for magnetic imaging of quantum materials and devices in various modes.**

Scanning superconducting quantum interference device microscopy (sSQUID) has become an important magnetic imaging technique to study superconductivity, magnetism and band topology in quantum materials and devices in reduced dimensions [1,2,3]. The direct flux sensitivity of sSQUID is a fundamental advantage over techniques relying on short-range magnetic interactions for magnetic sensitivity (e.g., MFM [4], spin-polarized STM [5], magnetic exchange-force microscopy [6], spin-polarized low energy electron microscopy [7], scanning NV microscopy [8]). It liberates sSQUID from stringent requirements on the cleanness of sample surface and greatly facilitates its application to a broad range of material systems, interfaces of heterostructures [9] and even device prototypes made of new materials [10,11].

In spite of the direct flux sensitivity of sSQUID, its spatial resolution is a serious disadvantage considering the ever-increasing need to resolve finer spin structures in a similarly broad range of materials and devices. Recently, a state-of-the-art spatial resolution of 50 nm has been achieved by the so-called SQUID-on-tip technique [12,13], where superconducting film composing the Josephson junctions was directly deposited on the tip of a fiber for scanning microscopy. Such a technique has been particularly successful in imaging vortex flow in superconductors [14] as well as exotic orders [15,16] in graphene. Nevertheless, more complex superconducting devices are generally difficult to fabricate on such a tip, hindering its functionality. Fabricating the SQUID on a silicon substrate (SQUID-on-chip, or SOC) allows multi-layered planar device structure which make a gradiometric SQUID loop [17] and the integration of additional superconducting loops becomes possible, such as field coils and modulation coils [18]. These coils enable flux-locked-loop operation [19] for magnetometry and simultaneous susceptometry imaging, which is very useful for mapping the superfluid density of two-dimensional superconductors [20]. However, the spatial resolution of SOC is fundamentally restricted by the distance from the pick-up coil and the chip edge during the tip fabrication. The distance of a few microns at best might be reached with mechanical polishing even under skillful hands [21]. On the other hand, partial etching of the tip area of the chip could reduce this distance but requires a very precise touch-down angle which is hard to achieve in practice [22].

In this work, we report the development of a batch fabrication process for SOC by taking advantage of the robustness of 3D nano-bridge junctions [23,24,25] so that deep silicon etching [26,27] could be utilized. This step generated a sharp edge around the pick-up loop which enabled scanning in close proximity to the sample surface. Combining such devices with high-precision control of sample-SOC distance, we achieved an imaging resolution of sub-micron spatial resolution in both magnetometry and susceptometry imaging of Nb chessboard test patterns of different sizes. In the 5-μm chessboard sample, we observed a strong magnetic signal at the edge of the square in the susceptometry images. We discuss the origin of this signal which will be helpful for future designs of SOC for susceptometry imaging.

The SOC integrates pick-up coils, modulation coils and field coils. These various functional coils allow us to perform simultaneous magnetometry, current imaging and susceptometry. We followed our previous designs in general [28,29] to prepare the devices. The whole process includes four layers of Nb wiring separated by $SiO_2$ as the dielectric, four steps of photolithography and one electron beam lithography (EBL) step. Josephson junctions were still made from 3D nano-bridge structures. The nano-bridge junctions not only are more robust during fabrications but also could work in a higher field than tri-layer junctions because of its much smaller area.

However, we implemented two other major steps in our fabrication process on 4-inch wafers in order to realize a close coupling between the pick-up coil and the sample. The first of them was that we used EBL technology to achieve nano-size pick-up coils instead of UV lithography. We fabricated the SOC with different inner diameters on the pick-up coils, $d$, from 4 μm to 150 nm [Fig. 1(a)]. The size of the field coils was 2 (4) μm inside (outside) diameter regardless of $d$. The second change was that we added a deep silicon etching at the last step to ensure minimum distance between the planar chip and the sample. We employed a robust photoresist AZ9260 of 8.2 μm thickness to protect the circuit area. Afterwards, we used the smooth Bosch deep etching [30] to completely etch through a 625-μm-thick silicon wafer. Since high temperature was necessary for this process, our 3D nano-bridge junctions, which were able to sustain the immense heat, had the unique advantage over tri-layer junctions to allow for such etching. The SOC after etching was suspended by two 100-micron supporting arms [Fig. 1(b)] to prevent breaking-off of chips. The edge of these devices was clean and the etched cross-section was smooth [Fig. 1(c)]. We were able to precisely control the etched edge directly to the outermost side of the field coil, the distance between the center of the pick-up coil and the etched edge was only 2.5 μm [Fig. 1(d)], much smaller than what was possible by mechanical polishing [21]. Taking into account the tilt angle of the chip, we estimated that the EBL technology combined with deep silicon etching provided the possibility to bring the center of the pick-up loop to within 250 nm of a flat sample surface.

We then demonstrated the electrical performance of the SOC in order to show they are still in working condition after the harsh etching process. We carried out current-voltage (I-V) as well as voltage-flux (V-Φ) characterizations at 4.2 K following standard procedure [28,29]. For SOC with a large pick-up coil, the characteristics were similar with those fabricated using previous technologies. Hence, we focus on the SOC with $d = 150$ nm. A typical I-V curve of them showed a critical current around 60 μA and normal state resistance of 4.4 Ω [Fig. 2(a)]. The V-Φ curves at various bias current exhibited a maximum modulation depth of 120 μV at 65 μA [Fig. 2(b)]. Such modulation depth was similar to that of the SOC fabricated by our previous process where deep etching was not employed. The flux noise of our SOC at 4.2 K with a SQUID array amplifier operating under flux-locked-loop was about 0.65 $\mu\Phi_0/\sqrt{Hz}$ at 100 Hz [Fig. 2(c)], slightly better than that of the previous generations without such technology [28]. These characterizations showed that full-depth silicon etching did not affect the performance of these devices, which allowed further utilization of these SOC for imaging with enhanced spatial resolution.

Besides the improvements on the SOC, the stability and precision of the scanning head is crucial for high resolution. To achieve that, we employed precise height control of the sample-SQUID distance by using the qPlus technique [31]. This technique has been developed for atomic-force microscopy and scanning tunneling microscopy and recently it was also used on SQUID-on-tip [32]. However, mechanical polishing employed on conventional SOC resulted in very blunt head that was difficult to take advantage of the qPlus technique. The deep silicon etching exposed a much sharper and straighter side-walls on the SOC [Fig. 1(c)] comparing with those polished mechanically [22], allowing a more sensitive height measurement using the tuning fork. We used torr steal to mount the SOC onto one of the prongs of a tuning fork [Fig. 3(a)] [33,34], whose other prong was fixed to a dither piezo. The dither mechanically vibrated under an oscillating driving voltage at the resonance frequency $f_0$ of the assembly [Fig. 3(b)]. The quality factor was reduced from a bare tuning fork due to the mass of the chip but was still enough for a precise height control. In order to minimize the detrimental effect of vibrations to high resolution imaging, we installed the scanning head in a closed-cycle cryostat with a remote valve. Vibration amplitudes of less than 0.15 nm was demonstrated in a similar system (bandwidth of 200 Hz, vertical direction) [35]. The vibration amplitude and the phase of the tuning fork changed dramatically when the tip of the nano-SQUID touched the sample surface [Fig. 3(c)]. The change appeared within a step size of 5 nm, demonstrating the overall precision and the stability of our scanning system with the SOC.

We made two checkerboard test samples in order to calibrate the spatial resolution of our SOC installed in this scanning system. The samples were checkerboard patterns of Nb islands deposited on $SiO_2$ substrates. The side length of the square Nb island was 4.4 μm and the interval was 5.4 μm [Fig. 4(a)]. Unlike the SQUID-on-tip technology, which is challenging to integrate additional coils and circuits with the SQUID, the SOC technology has a natural advantage at incorporating modulation coils and field coils which allows additional functionalities. The modulation coils enable flux-locked-loop [19,36] operations so that the measured flux signal is kept linear with the actual magnetic signal. The field coils enable local susceptibility measurements by passing a low frequency current $I_F$ through these coils. By demodulating the flux signal at this frequency, we obtain the susceptibility ($d\Phi/dI_F$) signal [28].

Since the spatial resolution of magnetometry and susceptometry are affected by quite different factors [22], we exam the images obtained on the test samples under these two modes separately. We first compared magnetometry images obtained by SOC's with $d = 150$ nm [Fig. 4(b)] and that of $d = 1$ μm [Fig. 4(c)]. The sample was cooled in ambient magnetic field. We did not observe any trapped vortices in either case because the small size of the Nb squares and their isolation from each other allowed the flux to be expelled into the film-free region. Since both chips were at a constant scanning height of 500 nm, it is evident that the image obtained by the small pick-up coil was much sharper than the one by the large coil. Furthermore, the amplitude of the flux signal from the small pick-up loop was about 30 times larger than that from the larger one. Since the flux signal from the same magnetic feature decreases quickly with the distance between the feature and the center of the pick-up loop, it showed that the $d = 150$ nm SOC was much closer to the sample surface than the $d = 1$ μm one. By taking line cuts through the edge of an island, we found the width of the steepest slope was around 800 nm. As this slope corresponds to 25% - 90% level of the maximum change on the step edge, we obtained a width of 2.17 μm for the $d = 1$ μm SOC using the same criteria [Fig. 4(d)]. The obtained width of these two SOC's are well consistent with the expected resolution given the geometry of the sample and the SOC's. Taking into account the linewidth of the pick-up coils (120 nm and 800 nm for the small and large probes, respectively), we found the tilt angles of the SOC's relative to the sample surface were 5.5 and 7.8 degrees, respectively.

Having shown much improved spatial resolution in the flux image, we analyzed the susceptibility image obtained by the same SOC with $d = 150$ nm on the 2-μm checkerboard structure [Fig. 4(e)]. Instead of a square pattern similar to the flux channel, susceptibility showed a more complex structure which only manifested the periodicity of the test pattern. Since the spatial resolution in susceptometry depends not only on the pick-up loop but also on the field coil as well as the response of the sample to the local field generated by the field coil, we performed detailed simulations by the 3D electromagnetic software Maxwell [37]. We used the exact same parameters in our simulation as the sample pattern, the pick-up loop and field coil, and simulated the response flux through the pick-up loop when a current was passed through the field coil as the SOC moved across over the edge of a square pattern [Fig. 4(f)]. Without using any free parameter, we found reasonable quantitative agreement between the measured and simulated susceptibility over this pattern. Employing the same criteria as the magnetometry (25% - 90% level of the maximum signal), we found a resolution of 1 μm in susceptometry by the SOC with $d = 150$ nm.

In contrast to the 2-μm checkerboard pattern, we observed unexpected features in the susceptibility image on the 5-μm pattern [Fig. 5(a)] using the SOC with $d = 150$ nm. Even though the diamagnetic signal from the Nb squares were more uniform and more similar in shape to the pattern, there appeared strong paramagnetic peaks at the edges of the squares. The width of the paramagnetic peak was around 800 nm [Fig. 5(b)], similar to the width of the steepest slope from magnetometry [Fig. 4(d)]. In contrast, such anomaly in susceptometry were absent when using a $d = 1$ μm SOC [Fig. 5(d) and (e)], which showed similar resolution and 25%-90% width as that of magnetometry [Fig. 4(c) and (d)]. Given the similar size of the square Nb island (4.4 μm) and the outside diameter of the field coil (4 μm), we speculate that the paramagnetic peak was a coincidental response due to the matching geometry when the pick-up loop was much smaller. When a small pick-up loop was located directly above the edge of a square, the size of the field coil was just large enough to induce Meissner response [38] from adjacent islands. Flux lines from such response were in the opposite direction of the diamagnetic response from the island beneath the pick-up coil. This island also repelled the flux lines from adjacent islands, leading to much enhanced flux density at the edge of the island and a strong paramagnetic signal. A more detailed study where more size variations of the pattern is investigated is necessary in the future to understand the origin of the paramagnetic peaks in susceptometry.

In conclusion, we have used electron beam lithography and deep silicon etching technologies to fabricate scanning SQUID gradiometric susceptometers based on 3D nano-bridge Josephson junctions on 4-inch wafers. The inside diameter of the pick-up loop of the SOC reached 150 nm. Combining this with the deep silicon etching allowed the center of the pick-up coil to be placed very close to a sharp edge of the chip so that a close coupling with the sample surface was realized. Placing such SOC on a tuning fork in a low vibration cryostat for precise height control, we achieved sub-micron spatial resolution both in magnetometry and susceptometry. We also showed that geometric effect can play a role in the susceptibility channel when the pattern size is very close to that of the field coil, which were valuable information for tuning the layout for further improvements in resolution. These results paved way for using the SOC to image quantum materials and devices with a much-enhanced spatial resolution.

**Acknowledgement**

YHW would like to acknowledge partial support by the Ministry of Science and Technology of China (2016YFA0302 and 2017YFA0303000), National Natural Science Foundation of China (Project 11827805) and Shanghai Municipal Science and Technology Major Project (Grant No. 2019SHZDZX01). LC would like to acknowledge partial support by the Frontier Science Key Programs of CAS (Grant No. QYZDY-SSW-JSC033), the National Natural Science Foundation of China (Grant No. 62071458), the Young Investigator program of the CAS (Grant No. 2016217) and the Strategic Priority Research program of CAS (Grant No. XDA18000000).

**Data Availability Statements**

The data that support the findings of this study are available from the corresponding authors upon reasonable request.

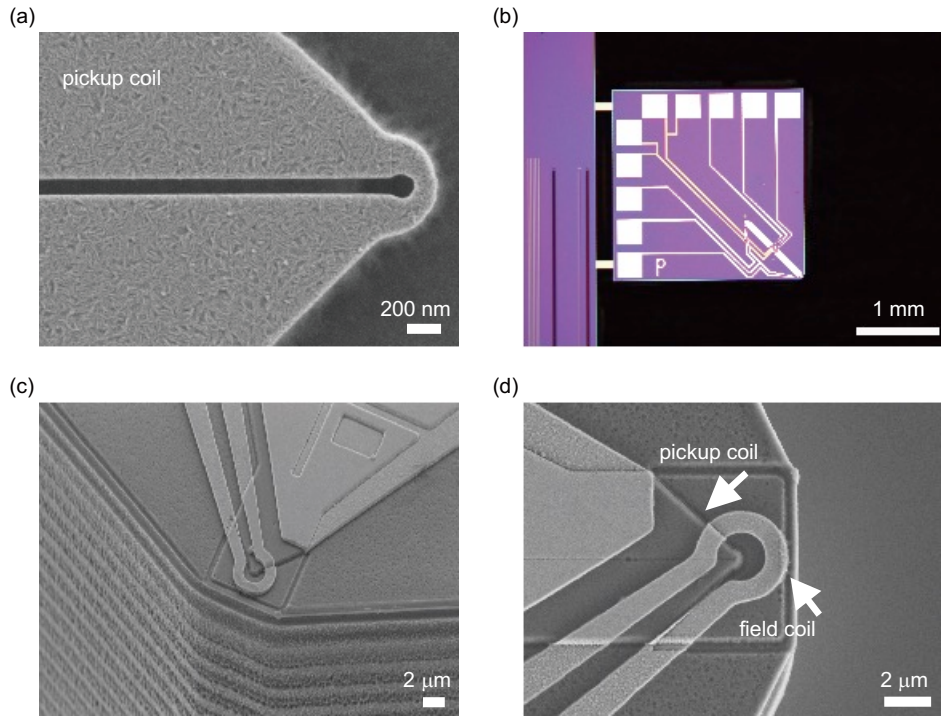

**FIG.1 Microscope and SEM images of a SQUID-on-chip device.** (a) is a zoom-in view of the pick-up coil with an inside diameter of 150 nm. (b) is an optical image of our SQUID-on-chip (SOC) device after deep etching. Two cantilever arms are visible. (c)-(d), Scanning electron microscopy images of the device. (c) shows the sharp edge after etching. (d) is a top view of the pick-up coil after deep silicon etching. The distance between the etched edge and the center of the pick-up coil is 2.5 μm.

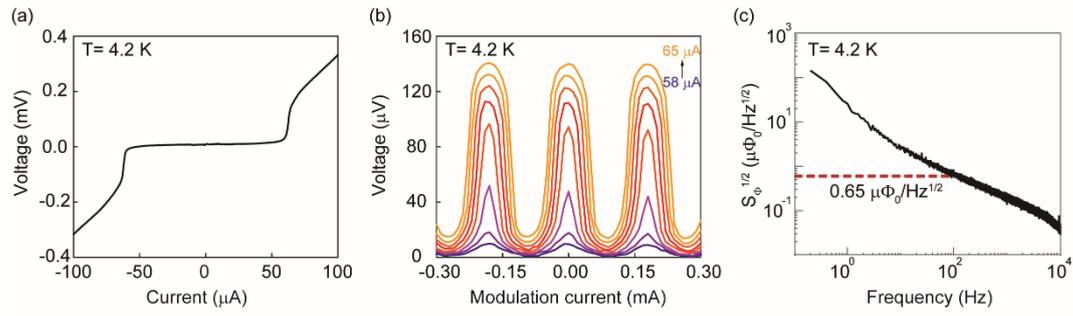

**FIG.2 Electrical transport characterization of a SQUID-on-chip device.** (a) Typical current-voltage characteristics of an SOC at 4.2 K. (b) SOC voltage modulation under different bias current at 4.2 K. The modulation current was applied through the modulation coils. (c) Typical flux noise spectrum at 4.2 K measured under a flux-locked-loop.

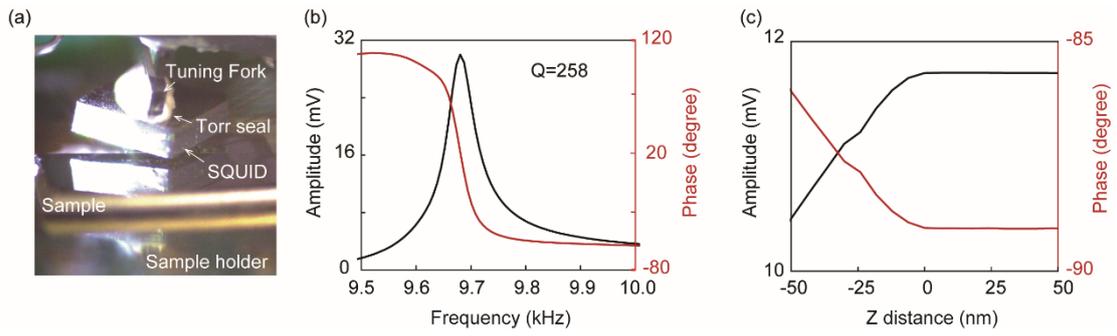

**FIG.3 Integration of SQUID-on-chip with a tuning fork for precise height control.** (a) An optical image of the SOC/tuning fork assembly as they were approaching the sample. (b) Amplitude (black) and phase (red) of the tuning fork with a mounted SOC as a function of the drive frequency at 1.6 K. (c) Changes in the amplitude and phase of the tuning fork as a function of the distance (Z) between the tip of the SOC and the sample. $Z = 0$ is the point where the two were touching.

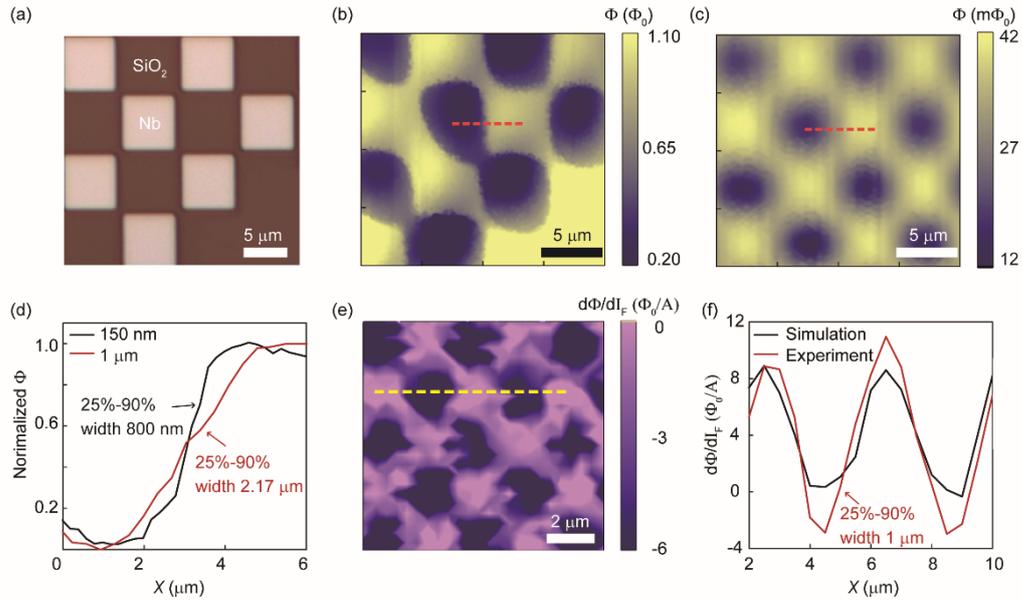

**FIG.4 Scanning SQUID flux and susceptibility images.** (a) is optical micrograph of a 5-μm checkboard Nb pattern. (b) and (c) are the flux images of (a), measured by an SOC with $d = 150$ nm (b), and one with $d = 1$ μm. (c), respectively. (d) Linecuts taken from the bright red dashed lines in (b) and (c), respectively. The 25% - 90% width corresponds to the steepest slope for the $d = 150$ nm SOC. The widths are 800 nm and 2.17 μm for the two pick-up loops, respectively. (e), Susceptibility image of a 2-μm checkboard Nb pattern obtained by an SOC with $d = 150$ nm (2 (4) μm inside (outside) diameter field coil). The red solid line in (f) is the linecut of the yellow dashed line in (e), and the black solid line is the simulation (see text).

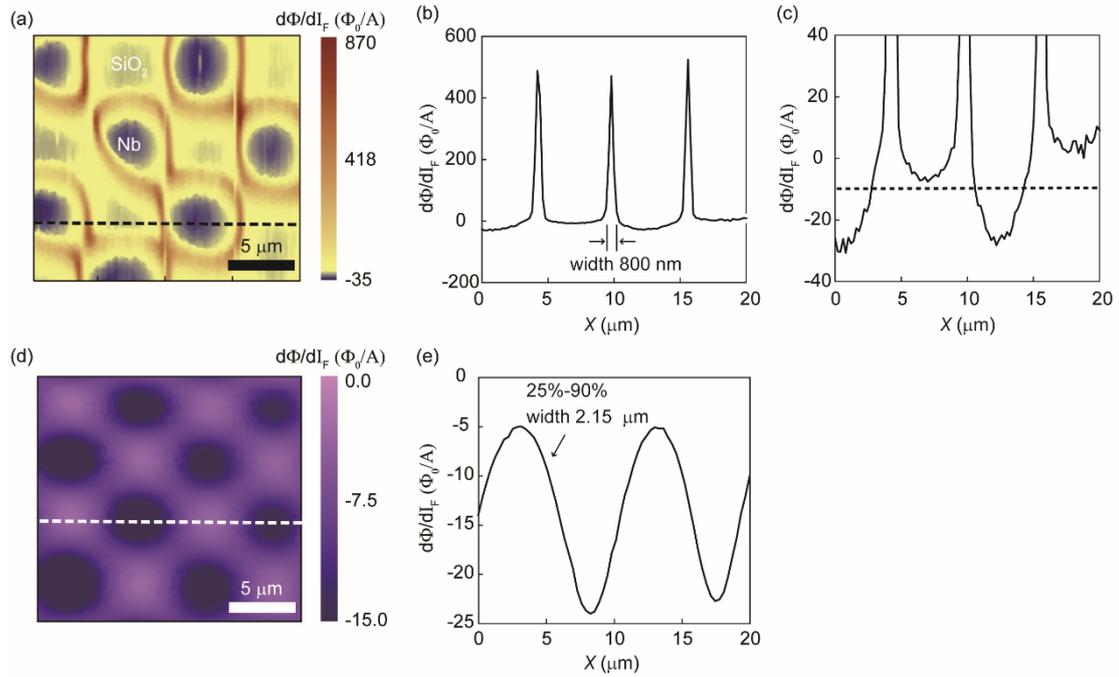

**FIG.5 Scanning SQUID susceptibility image of the 5-μm checkboard Nb pattern.** (a) is the susceptibility image by an SOC with $d = 150$ nm (2 (4) μm inside (outside) diameter field coil). (b) shows a linecut over the black dashed line in (a). (c) is a zoom-in of (b) showing the diamagnetic response in the middle of the Nb islands and paramagnetic peaks on the edges. (d) is the susceptibility image by an SOC with $d = 1$ μm. (e) shows a linecut over the white dashed line in (d).